\documentclass[a4paper]{article}

\usepackage{amsmath,amsmath,amsfonts,amssymb,color,bbm}

\usepackage{graphicx}

\def\Symp#1,#2,#3,#4.{\left[\left(\begin{array}{c}#1\\#2\end{array}\right),\left(\begin{array}{c}#3\\#4\end{array}\right)\right]}
\def\Vec#1,#2.{\left(\!\begin{array}{c}#1\\#2\end{array}\!\right)}
\newcommand{\beq}{\begin{equation}}
\newcommand{\eeq}{\end{equation}}
\newcommand{\beqa}{\begin{eqnarray}}
\newcommand{\eeqa}{\end{eqnarray}}
\newcommand{\ket} [1] {\vert #1 \rangle}
\newcommand{\bra} [1] {\langle #1 \vert}
\newcommand{\braket}[2]{\langle #1 | #2 \rangle}

\begin{document}
\title{\bf Anthropomorphic Quantum Darwinism as an explanation for Classicality.}
\date{}
\author{}
\maketitle
\vglue -1.8truecm
\centerline{\large Thomas Durt\footnote{TONA
Vrije Universiteit Brussel, Pleinlaan 2, B-1050 Brussels, Belgium. email:
thomdurt@vub.ac.be}}\bigskip\bigskip

\section*{Introduction.}

Presently, it is still an open question to know whether quantum mechanics is necessary
 in order to describe the way that our brain functions\footnote{It is even an open question to know whether the non-classical aspects of quantum mechanics play a fundamental role in biological processes at all. It is for instance an open question to know whether or not quantum coherence must be invoked in
order to explain intra-cellular processes. Nothing illustrates better the present situation than this
quote of Eisert and Wiseman\cite{[5]}:
''When you have excluded the trivial, whatever remains, however improbable, must
be a good topic for a debate"...}.

Nevertheless, quantum mechanics is astonishingly adequate if we want to describe the
 material world in which we live. It is therefore natural to assume that the way we think
  has something to do with quantum mechanics. After all, if our worldview faithfully reflects the 
  external world, it ought to reflect also its internal properties at the deepest level! For this reason, it is really interesting and important to reconsider epistemological questions in the light of the 
  most recent conceptual developments of the quantum theory. A key concept in these issues is the so-called quantum entanglement.

  The term entanglement was first
introduced by Schr\"odinger who described this as the characteristic trait of quantum mechanics,
``the one that enforces its entire departure from classical
lines of thought'' \cite{erwin}. Bell's inequalities \cite{key-73} show that when two systems are prepared in an entangled
state, the knowledge of the whole cannot be reduced to the knowledge of the parts, and that to
some extent the systems lose their individuality. It is only when systems are not entangled that they behave as separable systems\footnote{It can be shown that whenever two distant systems are
in an entangled (pure) state, there exist well-chosen observables such that the associated
correlations do not admit a local realist explanation, which is revealed by the violation of
well-chosen Bell's inequalities \cite{34}. In appendix (section \ref{bell}) we treat an example in depth and explictly derive Bell's inequalities that are violated in that special case.}. So, entanglement reintroduces holism and interdependence at a fundamental level\footnote{Holism is a rather vague concept that possesses several definitions, often mutually exclusive \cite{seevinck}. Here we mean that the quantum theory is holistic in the sense there can be a relevant difference in the whole without a difference in the parts. We provide in section \ref{fact2} an illustration of this property: the Bell states are different bipartite states for which the reduced local states (sections \ref{Schmidt} and \ref{SVN}) are the same. In this approach, entanglement, non-locality and non-separability are manifestations of holism and Quantum Weirdness, to be opposed in our view to the classical, Cartesian non-holistic approach in which the knowledge of the whole reduces to the knowledge of the parts.} and raises the following question: is it legitimate to believe in the Cartesian paradigm (the description of the whole reduces to the description of its parts), when we know that the overwhelming majority of quantum systems are entangled?

In order to tackle similar questions, that are related to the so-called measurement problem \cite{measprob}, Zurek and coworkers developed in the 
framework of the decoherence approach \cite{decoherence} the idea that maybe,
 if the world looks\footnote{The goal of the decoherence approach is to reconcile the first principles of the quantum theory, in particular the linearity of the quantum temporal evolution law with an objective description of the world. In the present context, when we write the world {\it looks} classical it means implicitly that we do not need an observer to let it look classical. As we explain in section 1, our approach is slightly different: we want to show that the world looks classical because our eyes are blind to Quantum Weirdness.} classical, this is because during the evolution, decoherence 
 selected in the external (supposedly quantum) world the islands of stability
  that correspond to the minimal quantum (Shannon-von Neumann) entropy \cite{reviewzurek,wikipedia}.

   In the present paper, we go a step further and make the hypothesis that these {\it classical islands} (environment induced or EIN superselected \cite{EIN}) would correspond to the structures that our brain naturally 
   recognizes and identifies, and this would explain why the way we think is classical. 
   
   In the first section we make precise in which aspects our approach coincides with and departs from the standard decoherence and Quantum Darwinist approaches and what is our motivation. 
   
   In the second section and in appendix, we explain the meaning of relevant concepts such as quantum entanglement, quantum bits, quantum non-locality and separability as well as Shannon-von Neumann entropy.  We also present a theorem that establishes that entanglement is the corollary of interaction (section\ref{i}) in the sense that when two systems interact, they get entangled in general. The classical situation for which  no entanglement is generated during the interaction is thus exceptional.
    
    In the third section we describe in more detail the environment induced (EIN) superselection rules approach and we
    apply it to the simple situation during which
     two quantum particles interact through a position-dependent potential, in
      the non-relativistic regime. We study then the classical islands that, according to the EIN selection rule, minimise the entropy of the reduced system, which means that they correspond to maximally deterministic (minimal uncertainty) states. They correspond to the classical, entanglement-free interaction regime. We show that the classical islands are in one to one
       correspondence with the three classical paradigms elaborated by physicists before quantum mechanics existed; these are the droplet or diluted particle model, the test-particle and the material point approximations (section \ref{iiv}). 
       
       The results presented in section \ref{ii} illustrate to which extent entanglement marks the departure from our classical preconceptions about the world, in agreement with Schr\"odinger's view \cite{erwin} according to which entanglement is the characteristic trait of quantum mechanics that enforces its entire departure from classical lines of thought.
       
They can be considered as a plausibility argument in favor of our main thesis according to which we are blind to Quantum Weirdness as a consequence of a long process of natural selection.
       
       \section{About the measurement problem and Quantum Darwinism.}
The measurement problem is related to the so-called objectification problem that in turn is intimately related to the Schr\"odinger cat paradox and, roughly, could be formulated as follows. Let us assume that we prepare two superpositions of macroscopically distinguishible quantum states (say a living cat state and a dead cat state). According to the quantum theory, whenever we perform a measurement that aims at revealing whether the cat is living or dead, one alternative is realized and the other one is discarded. Besides, if the state of the system obeys a unitary evolution law (like Schr\"odinger's equation) both alternatives survive and the system will remain in a superposition state. What is shown in the decoherence approach is that in good approximation the superposition becomes decoherent due to the interaction with the environment. What the decoherence approach doesn't prove is that one alternative is privileged\footnote{Roland Omn\`es for instance who is an active propagandist of the consistent history approach in which decoherence plays a central role, introduced in one of his books \cite{omnes} the Rule 5: {\it Physical reality is
unique. It evolves in time in such a way that,
when actual facts arise from identical antecedents,
they occur randomly and their probabilities
are those given by the theory.} In other words, Omn\`es must postulate that {\it Reality exists}; this is because he realized that the decoherence approach did not solve the measurement problem; in particular it could not solve the objectification puzzle so to say: if all alternative quantum possibilities, in a Schr\"odinger cat like experiment, do survive (with or without coherence that is not the point), then how is it possible that one of them is privileged and realized (actualized) in particularÉ? In his review of Omn\`es's book, W. Faris \cite{faris} wrote about Rule 5...{\it This statement by itself does not give a clear
picture of the mathematical formulation of actualization.
The intent may be that the notion of
ÒfactÓ is external to the theory, so that the rule
of actualization is merely a license to use consistent
logic to reason from present brute experience.
This is supported by the assertion:
ÒThe existence of actual facts can be added to
the theory from outside as a supplementary condition
issued from empirical observation.Ó A
dead cockroach is a fact; there is no more to it.
This is a long way from the ambitious goal of basing
everything on Hilbert space...}}. In order to show this, some extra-assumptions are required, for instance that the position is privileged (like in Bohm-de Broglie's interpretation) or that many worlds are present (here the world with a living cat and the world with a dead cat). In the Quantum Darwinist approach (developed in section \ref{QD}), it is assumed that somehow the objectification process will take place, and that it will occur in a privileged basis, the basis that diagonalizes the reduced density matrix of the system constituted by the quantum system under interest and the measuring apparatus. For instance if the murder of the cat is triggered by the measurement of a quantum two level system, the system under interest is this two level system, and the cat can be considered as a macroscopic amplifier or measuring apparatus.

Not every physicists is convinced (this is an euphemism) by Zurek's arguments \cite{janssen}, but Quantum Darwinism offers stimulating analogies with the biological world and with the Darwinian picture of the process of evolution. It are these features of Quantum Darwinism that stimulated the present work (more precisely, we were strongly stimulated by the postulated existence of a selection process that would ultimately lead to the selection of a preferred basis).

Another analogy between biological Darwinism and Quantum Darwinism is that in the latter the choice of the preferred basis is supposed to obey a principle of optimisation (similarly, in Darwin's approach, only the fittest species survive). At the same time many copies/duplicates of the preferred basis are supposed to be disseminated throughout the environment which is reminiscent of the reproduction mechanism of the surviving species.

According to us, Quantum Darwinism contains, like the many world or the Bohm-de Broglie interpretation a hidden extra-principle that goes beyond the standard principles of quantum mechanics, which is the Environment Induced Selection rule (see also section \ref{QD}). This rule tells us that the preferred basis is related to islands of classicality (section \ref{QD}) that minimise the Shannon-von Neumann entropy (in other words they minimise uncertainty, section \ref{SVN}) of the reduced state of the system constituted by the quantum system under interest and the measuring apparatus. We do not believe that this argument is very conclusive for the following reason: 

-the Shannon-von Neumann entropy is related to the distribution of the probability of occurence of events in the eigen-basis of the reduced state (density matrix), 

-but it does not make sense to talk about objective events and of their probabilitites as far as the measurement problem is not solved\footnote{All the interpretational problems of the quantum theory arise because the quantum theory is a probabilistic theory.}, 

-so that, in our view, in the Quantum Darwinist approach, one takes for granted from the beginning, in a subtle and implicit way, what one wants to prove at the end (the emergence of objective facts). 

In other words, we consider that in last resort the concept of entropy implicitly refers to an observer although the goal of the Quantum Darwinist approach is precisely to get rid of the dichotomy observed-observer, so that in our view the quantum measurement paradox is solved only at the price of introducing a logical loophole (circularity) somewhere in the reasoning.

It is not our goal to pursue in this direction: we have actually no clear idea about what is the right interpretation of quantum mechanics (if at least there exists such a thing, see e.g. the reference \cite{janssen} where the consistency of the decoherence approach is scrutinized and criticized in depth...). Nevertheless we shall exploit the idea according to which the preferred basis obeys an optimality principle: as has been succesfully shown by Zurek, classical islands are related to preferred bases (section \ref{QD}) in which the gathered information is reproducible, stable, maximally deterministic and insensitive to the disturbances of the environment.

Our next step is to seriously consider the hypothesis that humans (and possibly all sufficiently evolved living organisms, like cats for instance) select the preferred basis according to the Environment Induced  Selection rule (section \ref{QD}) because it is the most advantageous strategy in terms of the amount of stable and useful information that can be extracted from correlations, (the last idea is present in the Quantum Darwinist approach, the first one is a personal hypothesis).

At this point one should be very careful, we do not pretend in any way that the consciousness of the observer plays a role in the collapse process or anything like that\footnote{In agreement with Zurek, we do not believe that it is necessary to invoke consciousness in order to solve the measurement problem so to say to explain how objective, actual facts emerge from the quantum substrate. As we said before, we do not claim to solve the measurement problem in our approach. Our thesis is that we select preferentially classical correlations and that we are blind to Quantum Weirdness, which does never ever mean that consciousness is a quantum phenomenon or that we need consciousness in order to collapse the wave function. The correlations that we are talking about are correlations between events, clicks in detectors, reactions of sensors. Before correlations are treated by our nervous sytem, it is very likely that the decoherence process is achieved and that the effective collapse of the wave function already took place.}. No, our goal is rather to explain why 
our logic is classical although the logic of the quantum world is obviously non-classical. In other words we address the question to know why our representation of the world is classical although we live in a quantum world. Considered from this perspective, our answer is that maybe throughout eons of evolution, our ancestors (and their cousins, the ancestors of monkeys, cats, fishes and so on) became gradually blind to quantum interferences and quantum entanglement, because their manifestations (the associated correlations) were not useful from an informational point of view.

Our approach does not aim at solving the measurement problem (one still must postulate the existence of correlations, thus of probabilities between physical ''objective'' events, the existence of which is at the core of the measurement problem). Nevertheless, the novelty of our approach is to postulate that our sensorial system privileges data that are gathered in the preferred basis. So, what we retain from the Quantum Darwinist approach is that a preferred basis ''exists'', that it is the basis in which we measure the external world, and that this basis obeys an optimality principle that results from a (biological in this case) selection mechanism: the fittest is the best-informed, a reality that prevails in today's jungle, maybe more than ever.

In the section 3, we show that in a sense Quantum Weirdness is the rule, because entanglement, one of the most striking illustrations of Quantum Weirdness, can be considered to be the corollary of interaction. In simplified terms we can say that there is in general no interaction without creating entanglement. 

Now classical islands precisely correspond to the exceptional situations for which entanglement is negligibly small. This brings us to the section 4 where we can find the main novel contribution of the paper. In that section we aim at establishing the identity between classical islands and our cognitive representation of what elementary particles are. More precisely we study the regimes in which two particles interact (by a potential that depends solely on their distance) {\it without getting entangled}. These {\it ''entangled-free''} regimes are in one to one correspondence with the classical islands. We show that they also correspond to the models introduced by classical physicists before the advent of quantum physics in order to represent elementary particles such as the electron.  This observation is more a plausibility argument than a definitive proof. But it is our main argument in the favor of our personal interpretation of Quantum Darwinism.  

Metaphorically, our interpretation is that our ears are deaf to the quantum music because it would sound like a cacophony. Similarly we are blind to weird quantum optical effects because they would be like a very bad quality mirage to us (no stability, no reproducibility, no useful correlations to exploitÉ). After all, what we see is for instance not really what the detectors in our eyes perceive, there is still a huge work of data processing that occurs in the brain (in the case of vision, the corresponding hardware is not a negligible part of the brain!). We defend in this paper the thesis that we are blind to quantum effects not only because of the inadequacy of our receptors but mostly because of the treatment that we perform with the data gathered by them.

One could argue that at the level of our nervous system, quantum effects are so tiny that we should not be able perceive them. Our point of view is that it is not because they are tiny that we should not perceive them but rather that we do not perceive them because they are not useful\footnote{Particle physicists attempted to explain why organic molecules are chirally oriented in terms of the (tiny) energetic advantage that differentiates them from their mirror-molecule. In last resort this tiny difference of energy would be explained in terms of parity-violation by weak interactions \cite{chiral}.
We argue that the same effect, of amplification of tiny differences during millions of years, could explain the emergence of classical perceptions. If there was an informational advantage in exploiting non-classical quantum features like entanglement, it is likely that evolved organisms would be able to exploit this advantage. From this point of view, very primitive organisms, like bacteria would maybe be closer to the quantum world than we are, and it seems indeed, although no conclusive proof of this idea exists yet, that certain bacteria optimize their mechanism of harvesting light by exploiting the rich possibilities offered by the quantum superposition principle \cite{fleming}. Considered so, the receptors of those bacteria would exhibit quantum coherence, a fascinating hypothesis.\label{bact}}.
\section{Entanglement and Interaction.}
\subsection{The concept of entanglement.}

  In order to avoid unnecessary technicalities, and with the aim of addressing the paper to a large, interdisciplinary, audience, 
  we shall restrict ourselves in what follows to the simplest case: two systems $A$ and $B$ are prepared in a pure quantum state ${\bf \Psi}_{AB}$. 
  Then the state of the sytem is said to be factorisable (section \ref{fact2}) at time $t$ whenever it is the (tensor\footnote{The tensor product is represented by the symbol $\otimes$. It is the right mathematical operation that is needed when different Hilbert spaces are brought together, for instance the spaces associated to different particles. One can find a standard definition of Hilbert spaces and tensor products on the wikipedia website (\cite{hilbert},\cite{tensor}).}) product of pure quantum states of the subsystems $A$ and $B$, which means that the following constraint is satisfied: ${\bf \Psi}_{AB}( t)=
\psi_{A}( t)\otimes \psi_{B}( t)$. Otherwise the system is said to be entangled\footnote{ The characterization of entanglement can be generalized when the full state is not pure, or in the case of multipartite systems that are not bipartite but it is more complicate in this case \cite{bruss}.}. As we show in appendix (section \ref{fact2}), when they are in a non-entangled, factorisable, state, the two sub-systems $A$ and $B$ are statistically independent in the sense that 
the average values of any physical quantity associated to the subsystem $A(B)$ is the same that would be obtained if the system $A(B)$ was prepared in the state $\psi_{A}( t)(\psi_{B}( t))$. Moreover, there are no correlations at all between the subsystems and they can be considered to be independent.

The four so-called Bell two-qubit\footnote{A qubit is a 2-level quantum system. Examples of physical realisations of qubits are given in appendix (section \ref{qubit}).
} states \cite{bellstate} 
  are for instance entangled because as shown in section (\ref{fact2}) they do not factorize into a product of local qubit states.

 They are defined as follows \footnote{The state $\ket{B^1_{1}}$ is also known as the singlet spin state. }
:

  $\ket{ B^0_{0}}={1 \over \sqrt 2 }(\ket{ +}^A_{Z})\otimes \ket{ +}^B_{Z}+
  \ket{ -}^A_{Z})\otimes \ket{ -}^B_{Z})$

    $\ket{ B^0_{1}}={1 \over \sqrt 2 }(\ket{ -}^A_{Z})\otimes\ket{ +}^B_{Z}+
    \ket{ +}^A_{Z})\otimes\ket{ -}^B_{Z})$

    $\ket{B^1_{0}}={1 \over \sqrt 2 }(\ket{ +}^A_{Z})\otimes\ket{ +}^B_{Z}-
    \ket{ -}^A_{Z})\otimes\ket{ -}^B_{Z})$

    $\ket{B^1_{1}}={1 \over \sqrt 2 }(\ket{ +}^A_{Z})\otimes \ket{ -}^B_{Z}-
    \ket{ -}^A_{Z})\otimes\ket{ +}^B_{Z})$.

      Because the Bell states do not factorize,
   the local measurements in the distant regions $A$ and $B$ are not statistically independent,
    that is to say, the two qubits exhibit correlations. Moreover those correlations
    are essentially non-classical
   in the sense that they violate Bell's inequalities which is impossible for correlations derived from a local realistic model as we also show in appendix (\ref{bell}). 
   
\subsection{Entanglement and Interaction.\label{i}}
Entanglement between $A$ and $B$ is likely to occur whenever they interact \cite{quot} as shows the following property that we reproduce here without proof \cite{zeit}.

Let us consider two interacting quantum systems $A$ and $B$. We assume that the numbers of levels of the $A$ ($B$) system is finite and equal to $d_A$ ($d_b$)\footnote{The Hilbert
spaces associated to these systems are thus finite dimensional (of dimensions $d_A$ and $d_B$
respectively).}, that the wave-function of the full system is a pure state and obeys the Schr\"odinger equation: 
\begin{equation}\label{schrod}i \hbar \partial_t{\bf \Psi}_{AB}( t) =
H_{AB}( t){\bf \Psi}_{AB}( t)\end{equation}where $H_{AB}( t)$, the Hamiltonian of the system, is a self-adjoint
 operator, that we assume to be sufficiently regular in time. Then the following property is valid \cite{zeit}:

 {\it All the product states
remain product states during the interaction if and only if the full Hamiltonian can be
factorised as follows:
\begin{equation}\label{fact}H_{AB}(t)=H_{A}(t)\otimes I_{B}+I_{A}\otimes H_{B}(t)\end{equation}where
$H_{i}$ acts on the $i$th system only while $I_{j}$ is the identity operator on the $j$th system
($i,j=A,B$).}

 In simple words: there is no interaction without entanglement, which establishes that entanglement is very likely to occur; for instance, when we see light coming from a distant star, it is nearly certainly entangled with the atoms that it encountered underway. Entanglement can also be shown to be 
present in solid states, ferro-magnets and so on, and to play a very fundamental role in the macroscopic world, for instance during phase transitions \cite{[16]}.

\section{The decohererence program and the classical limit.\label{ii}}
\subsection{Environment induced superselection rules and classical islands.\label{QD}}

The EIN superselection approach was introduced by
Zurek in the framework of the decoherence approach \cite{wikipedia,decoherence,reviewzurek,EIN}. He postulated that the preferred basis obeys a principle of optimality that can be formulated in terms of the Shannon-von Neumann entropy\footnote{This quantity is defined in appendix (section \ref{SVN}). Entropy is a measure of the uncertainty of a probabilistic distribution. It is worth noting that very often entropy is also considered to be a measure of information, but that in our approach we consider that the useful information increases when entropy decreases. Indeed let us measure the degree of ''certainty'' or determinism assigned to a random variable by the positive quantity $C$, $C$= 1 - entropy; in other words certainty+uncertainty =1. We consider that when a distribution of probability is very uncertain (flat) it does not contain much useful information. On the contrary when a probability distribution is peaked it contains a high level of useful information. So when we use the word information in the paper we implicitly mean ''certain'' information or ''information useful for deterministic evolution'' (=1-entropy), a quantity that increases when entropy decreases, contrary to most commonly used conventions.}. More precisely, the EIN selection rule predicts that, once a system, its environment and their interaction are specified, the classical islands associated to the system are the states that diagonalize the reduced density matrix  of the system\footnote{In  appendix we explain that when the measure of entropy is the Shannon-von Neumann entropy of a density matrix $\rho$, $C$ is also a measure of the purity or coherence of $\rho$. Actually we have shown in ref.\cite{zeit} that when two coupled oscillators interact through the interaction Hamiltonian $H_{AB}=a^{\dagger}b+ab^{\dagger}$ (written in function of the creation and destruction phonon modes of the oscillators $A$ and $B$), states that remain factorisable throughout time are de facto eigenstates of $a$ ($b$), which means that they are so-called coherent oscillator states for which it can be shown that Heisenberg uncertainties are minimal.}. When the full state of the system-environment is pure, these states belong to the Schmidt basis (section \ref{Schmidt}) that diagonalizes the limit asymptotically reached in time by the system-apparatus bipartite state. When the environment can be considered as an apparatus, the classical islands define the so-called pointer basis\footnote{ {\it ``Pointer states can be defined as the ones which become
minimally entangled with the environment in the course of the evolution''}
(which means here temporal evolution described by Schr\"odinger's equation...),
quoted from Ref.\cite{abc}.}, also called sometimes the preferred basis. 

Roughly speaking, the EIN selection principle expresses that, during the evolution, the classical islands that belong to the prefered basis or pointer basis (the one that minimizes the Shannon-von Neumann entropy \cite{wikipedia,EIN} of the reduced system) are selected preferentially to any other basis. 
In the quantum Darwinist approach, the emergence of a classical world that obeys EINselection rules can be explained following two ways: 

A) these rules correspond to
maximal (Shannon-von Neumann) ''certainty'' or useful information\footnote{
The Shannon-von Neumann entropy of a reduced maximally entangled pure bipartite state, for instance of a Bell state is maximal (equal to 1) and the corresponding certainty (or useful information) minimal (0) while for a factorisable state the Shannon-von Neumann entropy of the reduced state is 0 and the certainty is equal to 1 (section \ref{SVN}).  As a consequence, factorisable states minimize the Shannon-von Neumann entropy of the reduced states. They correspond thus to classical islands, in the case that they are stabilized by the interaction with the environment.}; this fits to our own acceptance and interpretation of Zurek's Darwinism as we explained in the section 1: it is well 
plausible that our brain selects the features of the natural world that are maximally deterministic and minimally uncertain;

B) Zurek also invokes an argument of 
structural stability: superposition of states that would belong to such islands would 
be destroyed very quickly by the interaction with the environment which radiates irremediably the 
coherence (which is equal in virtue of our definitions to the certainty, so to say to 1-the Shannon-von Neumann entropy of the reduced density matrix of the system, see section \ref{SVN})) into the environment \cite{decoherence}. This process is called the decoherence process and is very effective. 

\subsection{A toy model for Quantum Darwinism: two interacting particles.\label{iiv}}

 We applied the Quantum Darwinist approach to a very simple situation during which the
system $A$ and the environment $B$ are two distinguishable particles and are described by a (pure) scalar wave function that obeys the non-relativistic
Schr\"odinger equation. We also assumed that their interaction potential $V_{AB}$ is an action a
distance that is time-independent and only depends on the distance between the particles, so that it is invariant under spatial translations (a Coulombian interaction
for instance). This is a standard text-book situation that was deeply studied, for instance in the framework of scattering theory. The systems $A$ and $B$ fulfill thus (in the non-relativistic regime) the following Schr\"odinger
equation:
\begin{displaymath}i \hbar \partial_t \Psi({\bf r}_{A} ,{\bf r}_{B} ,t) =
-({\hbar^2 \over 2m_A} \Delta_{A} + {\hbar^2 \over 2m_B}\Delta_{B})\Psi({\bf r}_{A}, {\bf
r}_{B} ,t)\end{displaymath}\begin{equation} + V_{AB} ({\bf r}_{A}-{\bf r}_{B})\Psi({\bf r}_{A} ,{\bf
r}_{B} ,t) \end{equation}
where $\Delta_{A(B)}$ is the Laplacian operator in the
$A(B)$ coordinates. 
Let us now consider that the system $A$ is the quantum system that interests us, and that the other system is its environment. 
Actually, the argument is symmetrical as we shall see so that this choice is a mere convention.
In order to identify the classical islands in this case, we must identify the states that exhibit maximal coherence or minimal Shannon-von Neumann entropy. We assume here that the full state is pure, which constitutes an oversimplification, because usually interaction with an environment destroys coherence. Nevertheless, as we shall show, one can get interesting insights even in this oversimplified situation.

Without entering into technical details that are presented in appendix (section \ref{SVN}), all we need to know at this level is that 
two systems in a pure state minimize the Shannon-von Neumann entropy when their state is factorisable or non-entangled.

Then, the classical islands 
correspond to the states that initially and during their interaction as well,  remain factorisable (more precisely in a pure factorisable state). 
This constraint can be shown \cite{zeit} to correspond to what is somewhat called in the litterature the mean field or effective field approximation, or Hartree approximation \cite{[7],gemmer}.
 In this regime,
particles behave as if they were discernable, and constituted of a dilute, continuous medium
distributed in space according to the quantum distribution $\big| \psi_{A(B)}(r_{A(B)},t)\big|^2$. Then, everything happens as if 
each particle ($A(B)$) ''felt'' the influence of the other particle as if it was diluted with a probability distribution equal to the 
quantum value $\big| \Psi({\bf r}_{B(A)})\big|^2$. 
It corresponds also to the concept of droplet or diluted particle\footnote{Actually, the diluted particle model corresponds to Schr\"odinger's own interpretation of the modulus square of the wave function, before Born's probabilistic interpretation was adopted by quantum physicists. The droplet picture is reminiscent of pre-quantum models of the electron that were developed by classical physicists such as Poincar\'e, Abraham, Laue, Langevin and others at the beginning of the 20th century. In this approach $|\psi|^2$ represents the charge density and as a consequence of Maxwell's laws, each particle {\it ''feels''} the Coulomb potential averaged on the distribution of the other particle.}
There are two interesting special cases: 

i) When the potential only depends on the relative position ${\bf r}_{rel}={\bf r}_{A}-{\bf r}_{B}$ $m_A<<m_B$, the initial state is factorisable 
and the $B$ particle is initially at rest and well localized,
 it can be shown that  the state remains factorisable in time and occupies thus a classical island. This corresponds to what is called the test-particle 
 regime (no feedback of $A$ onto $B$). For instance this is a good approximation of what happens in the hydrogen atom, where the electron is so light that it can be considered as a test particle.

ii) Another situation that is of physical interest is the situation of mutual scattering of two well localized 
wave packets  when we can neglect the quantum extension of
the interacting particles. This will occur when the interaction potential $V_{AB}$ is smooth enough and the
particles $A$ and $B$ are described by wave packets the extension of which is small in comparison to the
typical lenght of variation of the potential. It is well known that in this regime, when the de Broglie wave
lenghts of the wave packets are small enough, it is consistent to approximate quantum wave mechanics by
its geometrical limit, which is classical mechanics. For instance the quantum differential cross sections
converge in the limit of small wave-lenghts to the corresponding classical cross sections. Ehrenfest's theorem
also predicts that when we can neglect the quantum fluctuations, which is the case here, the average motions are
nearly classical and provide a good approximation to the behaviour of the full wave-packet so that we
can consider it to be a material point. Actually, in this regime, we can in good approximation replace the interaction potential
by the first order term of its Taylor development around the centers of the wave-packets associated to the
particles $A$ and $B$ so that the evolution equation is in good approximation separable into
the coordinates
${\bf r}_{A},{\bf r}_{B}$ \cite{zeit} and we have that, when 
$\Psi({\bf r}_{A} ,{\bf r}_{B} t=0) = \psi_A({\bf r}_{A}, t=0))\otimes \psi_B({\bf r}_{B} ,t=0)= \psi_A({\bf r}_{A}, t=0))\cdot \psi_B({\bf r}_{B} ,t=0)$, then, at
time
$t$,
$ \Psi({\bf r}_{A} ,{\bf r}_{B}, t)\approx \psi_A({\bf r}_{A} ,t)\cdot \psi_B({\bf r}_{B}, t)$
We shall discuss in the conclusion the relevance of this result.

   \section{ Conclusions and discussion.}
   
   The Quantum Darwinist approach sheds a new light on the emergence of classical logics and of our classical preconceptions about the world. The distinction between internal and external world, the Cartesian prejudice according to which the whole can be reduced to the sum of its parts and the appearance of preferred representation bases such as the position is seen here as the result of a very long evolution and would correspond to the most useful way of extracting stable and useful information from the quantum correlations.
   
   We conjectured in the present paper that our difficulties and resistances for conceiving ''entanglement'' are due to the fact that millions of years of evolution modelled our vision of the world, leading us to become blind to aspects of it that are not advantageous from the point of view of the acquisition of useful information.

 We showed that in a simplified situation (two particles that ''feel'' each other via an interaction potential), the EIN-selected classical islands  are
  regions of the Hilbert space where the 
 mean or effective field approximation (or Hartree approximation in the static case) is valid. In this regime, the interaction factorises into
the sum of two effective potentials that act separately on both particles, and express the average
influence due to the presence of the other particle. 

In that regime, it also makes sense to consider the particles, in accordance with classical logics, not as a whole but as separate objects. 

Our analysis confirms that our approach is well-founded in an undirect manner; indeed we show that the regime in which two particles interact without getting entangled possesses two extreme cases: the point particle regime (that corresponds to the classical material point description of matter) and the diluted matter approach (that corresponds to fluido-dynamics). The test-particle regime, where the heavy particle is treated like a material point, and the light particle as a diluted distribution, is intermediate between these two extreme cases\footnote{There are two interesting limits that are special cases of the Hartree regime (the test-particle and material points limit). The Hartree regime corresponds to the droplet model; the test-particle and material points limits correspond to the test-particle concept and to classical mechanics.
The Hartree regime is the most general regime in which entanglement is negligible.
The limit cases are obtained by neglecting the extension of one droplet, the one associated to the massive particle (the other particle appears then to behave as a test-particle) and of both particles (classical limit).
}. These ways of conceiving and describing matter are so deeply imbedded in our every-day representation of the physical world that it is nearly impossible to find an alternative representation for particles and atoms in our mental repertory. This explains according to us why it took such a long time for quantum physicists to realize the implications of the EPR paradox (30 years) and of the concept of entanglement. Even today, a well-trained quantum physicist can merely hope, at best, to reach a mathematical intuition of the phenomenon and entanglement remains a very counter-intuitive property.

It is interesting to note that somewhat similar conclusions could be drawned from the study of wave propagation, which is intimately related to propagation of information at the classical level, in other words of communication, another aspect of information\footnote{ Claude Shannon wrote hereabout {\it ''The fundamental problem of communication is that of reproducing at one
point either exactly or approximately a message selected at another point''} in his famous paper ÒA mathematical theory of communicationÓ \cite{shannon}.}.

A very interesting study was indeed performed at the beginning of the 20th century concerning the concept of dimension (see ref.\cite{huygens} and references therein): in a space-time of 1+1 or 3+1 dimensions wave propagation ruled by d'Alembert's equation obeys Huygens principle\footnote{Actually this is so in space-time of dimension $d+1$, where $d$ is an odd and positive integer.}, that can be translated into informational terms  \cite{ehrenfest}: the state of the image reproduces accurately the state of the source (after a time delay proportional to the distance between image and source). It is this property that allows us to obtain a fidel representation of the external world by using sensitive receptors such as our ears and our eyes. Also here one could invert the reasoning and assume that maybe the conventional 3+1 representation of space-time was privileged because it is informationally advantageous to do so. It could be that other physical phenomena that are characterised by other dimensions coexist but that we are blind to them simply because they are informationally deprived of interest and of sense\footnote{We are conscious that unfortunately this type of reasoning is not deprived of some degree of circularity, what is illustrated by the sentence {\it The world has 3 dimensions because we listen to music.} Therefore the best arguments that we can produce in their favor are plausibility arguments. This was precisely the scope of our paper (section \ref{ii}), in the framework of Quantum Darwinism.}.

Let us for instance excite a 2-dimensional vibrating membrane such as a drum skin by hitting at its centre. One can show that the signal at the edge at a given time is a convolution of the signal that was imposed at the centre of the skin in the past. The difference with what occurs in 1 and 3 dimensions is that the time-interval on which the convolution is taken has a non-negligible extension. Therefore correlations between the centre of the resonating membrane and the extremities are diluted and get lost, in close analogy to what happens during the decoherence process. 

It is extremely difficult for us to imagine the properties of a 4-dimensional space, which can be seen as a plausibility argument in the favor of a selection by our brain and sensors of a dimension (3) that optimizes the amount of useful information (in this case that optimizes efficient communication). A 2-dimensional space, a plane, is easy to visualize because, in our approach, it can be seen as a projection of the 3-dimensional space that we are supposed to live in.

Another promising direction of research was suggested by Nicolas Lori during the refereeing process of the paper. It concerns the possibility that certain ancient civilisations developed concepts similar to entanglement and inter-connectedness at an higher level than ours. This kind of research is outside of the scope of our paper, but it is worth noting that this observation could be brought in connection with the last part of footnote \ref{bact}.

Before ending the paper it is good to recall what the paper is about or rather what it is not about. We do not pretend to solve the measurement problem or the objectification problem. We do not pretend to settle definitively and unambiguously the question about where the collapse process (effective or not) would take place (in the brain or at the level of our physical sensors, or even before that in the external world, during a decoherence process). We claim that as the result of a natural selection process that privilegges the best-informed we became gradually blind to entanglement, but we are not categoric about where the blindness occurs: it could occur at the level of our sensors, or during the data treatment that occurs in the brain or at both levels simultaneously...The history of Quantum Mechanics has shown that one can survive without answering to all fundamental questions and paradoxes generated by the theory. In our view mystery and knowledge are to some extent complementary and undissociable.

\medskip                      
   \leftline{\large \bf Acknowledgment}
\medskip T.D. acknowledges support from the
ICT Impulse Program of the Brussels Capital Region (Project Cryptasc), the IUAP programme of the Belgian government, the grant V-18, and the Solvay Institutes for
 Physics and Chemistry. Thanks to Johan Van de Putte (Physics-VUB) for realizing the picture reproduced in figure 1. Thanks to the referees (Nicolas Lori, Alex Blin and anonymous) for useful comments and criticisms.

 \section{Appendix: Entanglement and non-local correlations.\label{ Noloc}}
  
        \subsection{Qubits.\label{qubit}}
A qubit consists of a two-level quantum system \cite{nielsen}.
This is the simplest conceivable quantum system, because a
one-level system would always remain in the same quantum state,
and it would be eternally static, which does not present any
interest from a physical point of view since physical theories are
mainly focused on transformations. The state of a two-level
quantum system is described, in the case of pure states, by a ray
of a two-dimensional Hilbert space. There exists several ways to
realize such systems, for instance, the qubit could be assigned
to degrees of freedom such as light polarization of a given
electro-magnetic mode,
 electronic, nucleic or atomic spin 1/2, energy of an orbital
 electron experimentally confined to a pair of energy levels, and so on.
In what follows, we shall most often assume that the qubit system
of interest is a spin-1/2 particle. Let us then denote $\ket{
+}_{Z}$ and $\ket{ -}_{Z}$ the spin up and down states relatively
to a conventional direction of reference $Z$. An arbitrary qubit
state can always be expressed as a superposition of the basis
states of the form
 $\alpha \ket{ +}_{Z}+\beta \ket{ -}_{Z}$ where $\alpha$ and $\beta$ are two normalized complex amplitudes: $\vert \alpha \vert^2+\vert
 \beta\vert^2=1$. In analogy with classical logic, we are free to
 associate with each of the basis states a conventional
  binary value yes-no or 0-1, for instance according to the assignment

   $\ket{ +}_{Z}\leftrightarrow  0 \leftrightarrow yes$,

   $\ket{ -}_{Z}\leftrightarrow  1 \leftrightarrow no$.

 Although in classical logic the value of a classical bit is either 0 or 1,
 a quantum bit or qubit can in general be prepared in a superposition state of the form
   $\alpha \ket{ 0}+\beta \ket{ 1}$ which offers more formal flexibility
   than in the classical case. Of course,
    during a measurement process, the outcomes are dichotomic. For instance,
    if we measure thanks to a Stern-Gerlach apparatus the projection of
    the spin along the $Z$ direction,
     the probabilities of the two possible outcomes spin up and down are
     respectively equal to $\vert \alpha \vert^2$ and
     $\vert\beta\vert^2$.  Despite of the fact that the distribution of
     the measurement outcomes is dichotomic,
      the evolution of the qubits, between the initial preparation and the
      final measurement,
      obeys the superposition principle.
      
\subsection{Two-qubit systems.}

   Let us now consider two spin 1/2 particles $A$ and $B$ 
that are localized in far away regions of space. 
 Let us measure with Stern-Gerlach devices their spin projection along $Z$,
 we get four possible outcomes:

 up$_{A}$-up$_{B}$,
 
 up$_{A}$-down$_{B}$,
 
 down$_{A}$-up$_{B}$
 
 and down$_{A}$-down$_{B}$.

    These outcomes correspond to the states
   
     $\ket{ +}^A_{Z}\otimes
   \ket{ +}^B_{Z}$,
   
    $\ket{ +}^A_{Z}\otimes
   \ket{ -}^B_{Z}$,
   
    $\ket{-}^A_{Z}\otimes
   \ket{ +}^B_{Z}$
   
    and $\ket{ -}^A_{Z}\otimes
   \ket{ -}^B_{Z}.$
   
   The most general two-qubit state is superposition of those 4 states:
 
    $\ket{ \Psi}=\alpha\ket{ +}^A_{Z}\otimes
   \ket{ +}^B_{Z}$+$\beta\ket{ +}^A_{Z}\otimes
   \ket{ -}^B_{Z}$+$\gamma\ket{-}^A_{Z}\otimes
   \ket{ +}^B_{Z}$+$\delta\ket{ -}^A_{Z}\otimes
   \ket{ -}^B_{Z}.$

\subsection{
Factorisable versus entangled  states.\label{fact2}}
A state that can be written as follows:

 $\ket{ \Psi}=(\alpha_{A}\ket{ +}^A_{Z}+\beta_{A}\ket{ -}^A_{Z})
\otimes
   (\alpha_{B}\ket{ +}^B_{Z}+\beta_{B}\ket{ -}^B_{Z})$
   
   is said to be factorisable. For such states, the outcomes of local measurements in the A and B region are independent. Indeed, local observables are of the type
  $O^A\otimes Id.^B$ ($ Id.^A\otimes O^B$) 
  so that

   $\bra{ i}^A_{Z}\otimes \bra{ j}^B_{Z}O^A. O^B\ket{ i}^A_{Z}\otimes
  \ket{ j}^B_{Z}$=$\bra{ i}^A_{Z}\otimes\bra{ j}^B_{Z}O^A\otimes O^B\ket{ i}^A_{Z}\otimes  \ket{ j}^B_{Z}=\bra{ i}^A_{Z}O^A\ket{ i}^A_{Z}\otimes
  \bra{ j}^B_{Z}O^B  \ket{ j}^B_{Z}$, 
   which means that outcomes of local measurements are statistically independent.

{\bf By definition: non-factorisable states are said to be entangled.}

The so-called Bell states \cite{bellstate} are massively used in the machinery of Quantum Information \cite{nielsen}, they provide a generic example of maximally entangled states.
They are in 1-1 correspondence with the well-known Pauli spin operators:

$\sigma_{0}=\ket{ +}_{Z}\bra{ +}_{Z}+\ket{-}_{Z}\bra{-}_{Z} \leftrightarrow 
\ket{ B^0_{0}}={1 \over \sqrt 2 }(\ket{ +}^A_{Z}\otimes\ket{ +}^B_{Z}+
  \ket{ -}^A_{Z}\otimes\ket{ -}^B_{Z}) $

$\sigma_{x}=\ket{ +}_{Z}\bra{ -}_{Z}+\ket{-}_{Z}\bra{ +}_{Z} 
\leftrightarrow 
 \ket{ B^0_{1}}={1 \over \sqrt 2 }(\ket{ +}^A_{Z}\otimes\ket{ -}^B_{Z}+
    \ket{ -}^A_{Z}\otimes\ket{ +}^B_{Z}) $

$\sigma_{y}=i\ket{ +}_{Z}\bra{-}_{Z}-i\ket{-}_{Z}\bra{ +}_{Z}
\leftrightarrow \ket{B^1_{0}}={1 \over \sqrt 2 }(\ket{ +}^A_{Z}\otimes
\ket{ -}^B_{Z}-
    \ket{ -}^A_{Z}\otimes\ket{ +}^B_{Z}) $

$\sigma_{z}=\ket{ +}_{Z}\bra{ +}_{Z}-\ket{-}_{Z}\bra{-}_{Z} \leftrightarrow
\ket{B^1_{0}}={1 \over \sqrt 2 }(\ket{ +}^A_{Z}\otimes\ket{ +}^B_{Z}-
    \ket{ -}^A_{Z}\otimes\ket{ -}^B_{Z}) $

    Bell states are not factorisable; for instance if $\ket{ B^0_{0}}$ 
  would factorize then
  
   $\alpha^A.\alpha^B$=$\beta^A.\beta^B$=$\sqrt{ 1\over 2} $ and
   $\alpha^A.\beta^B$=
   $\beta^A.\alpha^B$=$0 $; 
   
   Obviously such a system of equations has no solution because it implies that
   
   $\alpha^A.\alpha^B$.$\beta^A.\beta^B$=$\sqrt{ 1\over 2} $.$\sqrt{ 1\over 2} $=${ 1\over 2} $ and $\alpha^A.\beta^B$.$\beta^A.\alpha^B$=$0.0=0 $ so that finally $ 1/2=0 $, a logical contradiction, which shows the non-factorisable or entangled nature of the Bell states.

  To the contrary of factorisable states, when composite systems are prepared in entangled states, the distributions of outcomes observed during local observations are no longer statistically independent.

For instance, let us assume that the qubit systems $A$ and $B$ are prepared in the Bell state $\ket{ B^0_{0}}$ and let us measure the spin projection along $
   \vec n_{A}$ in the region $A$ and the spin projection along $
   \vec n_{B}$ in the region $B$ (with $n^{A/B}_{x}= sin\theta^{A/B}$,
  $n^{A/B}_{y}=0$, $n^{A/B}_{z}=cos\theta^{A/B}$). In order to evaluate the corresponding distribution of outcomes, we can make use of the spinorial transformation law
  
  $\ket{+}_{\vec{ n}}=\cos{\theta\over 2}e^{-i\phi\over 2}\ket{
+}_{Z}+
\sin{\theta\over 2}e^{+i\phi\over 2} \ket{ -}_{Z}$ and $
\ket{-}_{\vec{ n}}=-\sin{\theta\over 2}e^{-i\phi\over 2}\ket{
+}_{Z} +\cos{\theta\over 2}e^{+i\phi\over 2} \ket{
-}_{Z}$, where $\theta$ and $\phi$ are the polar angles associated to the direction $\vec n$ (here $\phi_A=\phi_B=0)$, so that the Bell state $\ket{ B^0_{0}}$ transforms as follows:
  
    \beqa \ket{ B^0_{0}}=\sqrt{ 1\over 2}(
 cos{(\theta_{A}-\theta_{B})\over 2}\ket{ +}^A_{\vec{ n}}\otimes\ket{ +}^B_{\vec{ n}} \nonumber\\
 -sin{(\theta_{A}-\theta_{B})\over 2}\ket{ +}^A_{\vec{ n}}\otimes\ket{ -}^B_{\vec{ n}}\nonumber \\
 +sin{(\theta_{A}-\theta_{B})\over 2}\ket{ -}^A_{\vec{ n}}\otimes\ket{ +}^B_{\vec{ n}}\nonumber \\
 +cos{(\theta_{A}-\theta_{B})\over 2}\ket{ -}^A_{\vec{ n}}\otimes\ket{ -}^B_{\vec{ n}}).\nonumber
 \label{A-B}\eeqa

 Making use of Born's transition rule, the probability that after the preparation of the Bell state $\ket{ B^0_{0}}$ the outcomes of the 
spin measurements in $A$ and $B$ 
are found to be up-up is equal to
  $\vert \bra{ B^0_{0}}(\ket{ +}^A_{\vec{ n}}\otimes\ket{ +}^B_{\vec{ n}})\vert^2$ so to say to
   ${ 1\over 2}cos^2{(\theta_{A}-\theta_{B})\over 2}$.
  Similarly the probability of $(up_{A},down_{B})$ is ${ 1\over 2}
 sin^2{(\theta_{A}-\theta_{B})\over 2}$, the probability of $(down_{A},up_{B})$ is ${ 1\over 2}sin^2{(\theta_{A}-\theta_{B})
 \over 2}$, and the probability of $(down_{A},down_{B})$ is ${ 1\over 2}cos^2{(\theta_{A}-\theta_{B})\over 2}$.
  
  In particular, when local quantization axes are parallel:
 ($\theta_{A}-\theta_{B}$=0), we get perfect correlations:

$P(up_{A},up_{B})=P(down_{A},down_{B})=1/2$

$P(down_{A},up_{B})=P(up_{A},down_{B})=0.$
 
 Obviously there is no longer statistical independence; otherwise we would get 

$P(up_{A},up_{B})$.$P(down_{A},down_{B}) $=$1/2.1/2$=$P(down_{A},up_{B})$.$P(up_{A},down_{B})=0.0$ so that finally $1/4=0$, a logical contradiction.

We shall show in the next section that correlations exhibited by entangled systems have no classical counterpart.

 \subsection{Bell's inequalities.\label{bell}}        
   
 Let us consider a situation \`a la Bell \cite{key-73} during which a pair of qubits is
   prepared in the entangled state $\ket{ B^0_{0}}$. A Stern-Gerlach measurement
   is performed along the direction $ \vec{n_{A}}$ in the region $A$ and another Stern-Gerlach measurement is performed simultaneously
     along the direction $ \vec{ n_{B}} $ in the distant region $B$ (see fig.1\footnote{This is a schematic representation, in the case that the the particles are emitted along opposite directions along the $Y$ axis, with the source in-between the regions $A$ and $B$, that we did not represent on the picture in order not to overload the representation.}).

  As we noted before, whenever the directions
  $ \vec n_{A} $ and $ \vec n_{B} $ are parallel, the outcomes are
  maximally correlated in the sense that the probability that the outcomes observed in the
  $A$ and $B$ regions are different is equal to 0. We thus face a situation in
  which we could in principle predict the outcome that is observed during
  a measurement in one of the regions, simply by performing
  the same measurement in the other region.
  By itself, this situation has nothing special: it could occur
  that making use of maximal correlations, a classical observer can
  infer with absolute certainty the validity of certain properties
  without testing them directly.

  What is puzzling is that what quantum mechanics says about the value of
  the local spin is that it could be
  up with probability 50 percent and down with probability 50 percent,
  and that the formalism gives the feeling that the transition does not occur before the
  measurement process occurs.

  It would mean that we ''create'' the value of the spin only when we look at it, a rather counter-intuitive result. Intuitively,
  we are likely to think that our observation only reveals pre-existing properties,
  since this is so, as far as we know, at the classical level.
  This is particularly obvious when $A$ and $B$ are separated by
  a space-like distance because, even if one accepts that the measurement
  influences the system under observation, it is difficult to understand
  how this influence would occur instantaneously, thus faster than the speed of light.
  Our classical intuition thus suggests the existence of pre-existing properties-
  and this is essentially the reasoning held
  by Einstein, Podolski and Rosen \cite{key-74} in 1935-
  that the value of the spin pre-existed before the measurement process.
  If this is so, we are led to infer
  from the EPR reasoning that some deterministic ''element of reality'' is present but
  then this information is hidden and lost at the level of the quantum formalism,
  because the prediction of quantum mechanics is simply
  that each possible outcome (up or down) is observed with probability fifty percent,
  a purely indeterministic prediction.

  The reasoning of EPR did not go much further than this; their final remark was that,
  given that a hidden determinism is present and that such a
  hidden determinism is not present at the level of the quantum formalism,
  the quantum theory is not complete and ought to be completed by a (hidden)
  deterministic or local realistic hidden variable theory (HVT).

  In 1965, John Bell went further \cite{key-73,bellspeakable} and showed that the existence of hidden determinism is
  incompatible with the predictions of quantum mechanics. We shall reproduce here the
  essence of his reasoning, and, following Ref.\cite{squires} (see also references about Pseudo-telepathy in ref.\cite{swap}), we shall enhance the dramatic
           character of the result by assuming that two persons, Alice and Bob,
           make use of the results of the Stern-Gerlach measurements on the qubits $A$ and $B$
           respectively in order to simulate a ''telepathic''
        relationship. During this ''performance'', Alice and Bob are located in far away regions; for instance,
        Alice is on earth while Bob is located inside an interplanetary rocket, more or less one
         lighthour away from earth. Both are kept in isolated, high-security, cells and are not allowed to communicate with each other.
         Every hour, a guardian $A$ enters Alice's cell and asks a
         question that is chosen at random among three possible questions $\alpha$, $\beta$
          and $\gamma$.
          For instance the questions could be:

         $\alpha$: Are you thirsty?

         $\beta $: Are you tired?

         $\gamma$: Are you happy?

         Exactly at the same time, (which means in this precise case simultaneously
         relatively to (an inertial frame comoving with the center of mass of) the solar
         system), a guardian $B$ enters Bob's cell and asks a
         question that is chosen at random, and independently on the choice performed by the guardian $A$,
         among the selection $\alpha$, $\beta$ and $\gamma$.
         We also assume that the experiment is repeated many many times, hour after hour, in order to establish a relevant statistics of
         the correlations between Alice and Bob's answers.

         Another rule of the game is that each time they are presented with a question,
         Alice and Bob must answer at once and have two possible answers: Yes and No.

         Let us now assume that Alice and Bob make use of a quantum device in order to
         answer the questions: they share a pair of qubit states prepared in the
          entangled state $\ket{ B^0_{0}}$, and in order to answer the questions $\alpha$,
           $\beta$, or $\gamma$,
           they measure thanks to a Stern-Gerlach device the spin of the qubit in their possession along
         the directions $ \theta_{\alpha}=0, \phi_{\alpha}=0$;
          $ \theta_{\beta}=2 \pi/3, \phi_{\beta}=0$; or
          $ \theta_{\gamma}=4 \pi/3, \phi_{\gamma}=0$.

          Because of the perfect correlations exhibited by the Bell state $\ket{ B^0_{0}}$, whenever
           Alice and Bob are asked simultaneously the same question they
           will provide exactly the same answer, which tends to simulate a
           telepathic communication between them.

           The first reaction, confronted with such a situation, would be to make the rational
           hypothesis according to which Alice and Bob possibly cheat by sharing a same
           list on which they have written in advance all possible answers to the three questions, at all times. Before they answer, they would consult the list and answer accordingly.
           This is nothing else, in the present context, than EPR's hypothesis.

           John Bell \cite{key-73} went further by showing that if such a list existed, the correlations
            ought to obey certain constraints (inequalities), and that those inequalities are
             violated by the quantum correlations, which renders impossible the existence of
              a list of pre-existing outcomes; in other words, the violation of Bell's inequalities denies
               the possibility of explaining quantum correlations by a local realistic HVT.

           In the present case, it is easy to derive such an inequality,
           following the approach of Ref.\cite{gold} and making use of a property
           that was baptised by mathematicians under the name of pigeonhole's principle. The
           idea is simple: let us assume that three pigeons plan to spend the night in the holes of
           a cliff; if there are only two holes then certainly at least two pigeons will have to sleep
            in the same hole. This kind of reasoning is used for instance to show that within a
            population of $10^6$ persons, at least two persons will have exactly the same number
             of hair.

 Now, there are three questions and two answers so that, in virtue of the pigeonhole principle, two questions will share the
 same answer and we can write the equality:
  \beq P(\alpha_{A}=\beta_{B}\vee \beta_{A}=\gamma_{B}  \vee \gamma_{A}=\alpha_{B} )=1 ,\eeq
 \noindent where $\vee$ expresses the logical disjunction (''or'') and the equality symbolically means that two questions have the same answer, for instance $\alpha_{A}=\beta_{B}$ means that the measured values of the spins along the directions $\alpha_{A}$ and $\beta_{B}$ are the same (either both up or both down).
          Now, it is well-known that the probability of the disjunction of two or
          more properties is less than or equal to the sum of their probabilities
          so that we can write the Bell-like inequality
           \beq P(\alpha_{A}=\beta_{B})+P( \beta_{A}=\gamma_{B} )+P( \gamma_{A}=\alpha_{B} )\geq 1       .\eeq

           This inequality is violated by quantum correlations because $P(\alpha_{A}=\beta_{B})+P( \beta_{A}=\gamma_{B} )+
           P( \gamma_{A}=\alpha_{B} )$=
           $P(\alpha_{A}=\beta_{B}=up)+P( \beta_{A}=\gamma_{B}=up )+P( \gamma_{A}=\alpha_{B}=up )$+
           $P(\alpha_{A}=\beta_{B}=down)+P( \beta_{A}=\gamma_{B}=down )$+$P( \gamma_{A}=\alpha_{B}=down )$
           =$6.{1\over 2}cos^2(\pi/3)=3/4$. It is not true that
           $3/4\geq 1$ and the inequality is violated.

           Of course there are other logical explanations of the correlations: it could be that
           Alice and Bob secretly communicate, but as their distance is of the order of one
           light-hour and that they must answer at once (within say one second),
           they have to communicate more or less 3600 times faster than light.

           Actually a similar situation was experimentally implemented in the surroundings of Geneva: instead of one
            light-hour the distance was ten kilometers and instead of one second the duration was of the order of 10 picoseconds.
             This imposes the experimental limits according to which,
              if Alice and Bob cheat and communicate in secret, they must do it
              $7.10^6$ times faster \cite{key-72,london} than
              light. In the
              literature, this (hypothetical) ¥phenomenon is called non-locality, and is reminiscent of the mysterious Newtonian action-at-a-distance.

\subsection{About the Shannon-von Neumann entropy and the bi-orthogonal (Schmidt) decomposition.\label{Schmidt}}

One can show \cite{cosmos,physicalia}) that when a bipartite system is prepared in the
 (pure) state
  $\ket{ \Psi}^{AB}=\sum_{i,j=0}^{d-1}\alpha_{ij}\ket{ i}^A\otimes \ket{ j}^B$ (where $\ket{
i}^A$ and $\ket{ j}^B$ are states from orthonormalized reference
bases) everything happens ''from A's point of view'' as if he had
prepared his system in the state described by the effective or
reduced density matrix
$\rho^A=\sum_{i,i'=0}^{d-1}\sum_{j=0}^{d-1}\alpha_{ij}^*\alpha_{i'j}
\ket{ i'}^A\bra{ i}^A.$

Now, it can be shown that the reduced density matrix has all the
properties of density matrices (its trace is equal to one, it is
a self-adjoint operator with a positive spectrum) so that we can
find at least one basis in which it gets diagonalized, that we
shall  label by tilde indices ($\ket{\tilde  i}^A$): $\ket{
\Psi}^{AB} =\sum_{i,j=0}^{d-1}\tilde {\alpha}_{ij}\ket{ \tilde
i}^A\otimes \ket{ j}^B$$=\sum_{i=0}^{d-1}\ket{ \tilde
i}^A\otimes(\sum_{j=0}\tilde {\alpha}_{ij} \ket{ j}^B)=\sum_{i=0}
^{d-1}\alpha_{i}\ket{ \tilde i}^A\otimes\ket{ \tilde i}^B$ where
we introduced the notation $\alpha_{i}\ket{ \tilde
i}^B=\sum_{j=0}\tilde {\alpha}_{ij} \ket{ j}^B$, with
                  $\alpha_{i}$ a normalization factor.
The states $\ket{ \tilde i}^B$ are necessarily orthogonal,
otherwise they would generate off-diagonal interference terms in
the reduced density matrix, which may not occur because the basis
states $\ket{\tilde  i}^A$ diagonalize the reduced density matrix
of A's subsystem.

This proves that we can always write a bipartite pure state
$\ket{ \Psi}^{AB}$ in the so-called bi-orthogonal form
\cite{peres}:

$\ket{ \Psi}^{AB} =\sum_{i=0}
                ^{d-1}\alpha_{i}\ket{ \tilde i}^A\otimes\ket{ \tilde i}^B$, where the states
                 $\ket{ \tilde i}^A$ ($^B$) are orthonormalized. This form is called
bi-orthogonal because the matrix ${\alpha}_{ij}$ is in general a
non-diagonal matrix. It is only when the full state is expressed
in the product of the bases
                 composed by the
                 states that diagonalize the reduced density matrices that the amplitudes-matrix
                  becomes diagonal: $\tilde {\alpha}_{ij}= \alpha_{i}\delta_{i,j} $.

When the state is expressed in its biorthogonal form, it is easy
to analyze the degree of entanglement of the two subsystems. One
can quantitatively estimate the degree of entanglement by
counting the number of coefficients $\alpha_{i}$ that differ from
zero (this is called the Schmidt number). The Shannon entropy of
the probability distribution $\vert \alpha_{i}\vert^2$, which is
also equal to the Shannon-von Neumann entropy of the reduced density matrix of A or B's subsystem (section \ref{SVN}), provides a more
quantitatively precise parameter in order to estimate their
degree of entanglement (it is equal to 0 for factorizable states,
in which case the biorthogonal decomposition contains only one
factor, and equal to 1 when the state is maximally entangled so
to say when $\vert \alpha_{i}\vert^2=1/d, \forall i$). It is easy
to check that Bell states are maximally entangled, which
corresponds to a density matrix proportional to the identity
operator. Such a density matrix is said to be totally incoherent
because in all conceivable interference experiments it will
always exhibit a flat interference pattern (of ''visibility''
equal to 0).

Actually the purity or coherence which is equal to 1-the Shannon-von Neumann entropy of a reduced density matrix measures the
degree of anisotropy exhibited by the corresponding state in the
Hilbert space. When a state is pure it determines a preferred
orientation (or ray) in the Hilbert space. An incoherent state
is, from this point of view, totally isotropic and indeed the
probability of transition of such a state to any pure state is
constant and equal to 1/$d$. This explains why such states always
exhibit totally flat interference patterns.

       \subsection{ Entanglement, non-separability and loss of identity.\label{SVN}}

Another aspect of entanglement is its ''fusional'' nature which we consider to be a manifestation of quantum holism. Bell's
analysis of the nature of quantum correlations shows that, in
contradiction with the Cartesian paradigm, when two systems are
prepared in an entangled state, the knowledge of the whole cannot
be reduced to the knowledge of the parts, and that to some extent
the systems lose their individuality.   It is only when their
joint wave-function is factorizable that they are
separable\footnote{ As we mentioned before, whenever two distant systems are in
an entangled (pure) state, it has been shown \cite{34} that there
exist well-chosen observables such that the associated
correlations do not admit a local realist's explanation, which is
revealed by the violation of well-chosen Bell's inequalities.}. A
very interesting uncertainty (complementarity \cite{walther})
relation characterizes the entanglement of a pair of quantum
systems prepared in a pure state: the entanglement of the whole
system (that measures its degree of inseparability) is
equal to the Shannon-von Neumann entropy of the
reduced system, which is also a ngeative measure of the coherence of the system. This relation is expressed by the equality
$E(A-B)=1-C(A)=1-C(B)$, where $C(A(B))=1+Tr(\rho^{A(B)}
log_d\rho^{A(B)})$, the Shannon-von Neumann coherence of the
subsystem $A$ ($B$) which measures the degree of purity or
coherence of their reduced state, as well as the degree of certainty associated to this state, while $E(A-B)$ is, in the case
of pure states, a ''good'' measure of the entanglement between the subsystems $A$ and $B$. In simple terms, the
irreducibility of the whole to the parts (or entanglement between
them) increases when the Shannon-von Neumann measure of the ''certainty'' of the
parts (or their ''purity'' or degree of ''coherence'') decreases and vice versa.    For instance, when the
state of the full system is pure and factorizable, their
entanglement is equal to 0 and the reduced system is a pure state
with a minimal Shannon-von Neumann entropy equal to 0 (maximal coherence equal to 1). When
the full system is prepared in a Bell state, their entanglement
is maximal and equal to 1 and the reduced system is a totally
incoherent density matrix proportional to the identity, with a
maximal Shannon-von Neumann entropy equal to 1 (minimal coherence equal to 0). This
complementarity relation can  be generalized when the full state
is not pure, but the situation is more involved in this case
\cite{shimony,10}, among others because there exists no simple measure of the entanglement of two subsystems when the
system is prepared in a mixed state \cite{10}.

If metaphorically we transfer this idea to human relationships, it could be
translated (in a very free way, because there is no direct
counterpart for the concept of quantum purity at the human level)
by something like the ''fusional nature of entanglement'': when
$A$ and $B$ are strongly entangled, they lose to some extent
their individuality. We mean that the coherence of the parts
decreases, and the coherence or purity is seen here as a measure
of the independence (singularity) relatively to the rest of the
world. This fusional nature is contagious to some extent (the
friends of my friends are my friends) because it can be shown that
two systems $C$ and $D$ can become entangled although they never
interacted directly and remain spatially separated, provided they
get entangled (through interaction for instance \cite{zeit,quot})
with subsystems that are entangled (this is called entanglement
swapping-see e.g. ref.\cite{swap} and references therein). For
instance, regions that are separated by huge distances in the
galaxy \cite{braun} can be shown to become entangled because they
both interact with  the cosmic background radiation which
presents a non-negligible degree of spatial entanglement. Another related property is monogamy: fusional relations are often exclusive, which possesses a quantum counterpart the so-called quantum monogamy \cite{monogamy,cosmos}.

\end{document}